\documentclass[reprint, superscriptaddress, amsmath,amssymb, prb, showpacs, altaffilsymbol]{revtex4-2}

\usepackage{graphicx}
\usepackage{epsfig}
\usepackage{dcolumn}
\usepackage{bm}
\usepackage{xcolor}
\usepackage{braket}
\usepackage{amsmath}
\usepackage{amssymb}
\usepackage{mathtools}
\usepackage{breqn}
\usepackage{float}
\usepackage{makecell}
\usepackage{natbib}
\usepackage{ulem}

\usepackage{gensymb}

\makeatletter
\let\cat@comma@active\@empty
\makeatother

\begin{document}

\preprint{APS/123-QED}

\title{Multi-Band Superconductivity in Strongly Hybridized 1T'-WTe$_2$/NbSe$_2$ Heterostructures}
   

\preprint{APS/123-QED}

\author{Wei Tao}
\author{Zheng Jue Tong}
\affiliation{Division of Physics and Applied Physics, School of Physical and Mathematical Sciences, Nanyang Technological University, Singapore 637371, Singapore}

\author{Anirban Das}
\affiliation{Department of Physics, Indian Institute Of Technology Madras, Chennai, Tamil Nadu 600036, India}
\affiliation{Computational Materials Science Group
IIT Madras Chennai, Tamil Nadu 600036, India}

\author{Duc-Quan Ho}
\affiliation{Division of Physics and Applied Physics, School of Physical and Mathematical Sciences, Nanyang Technological University, Singapore 637371, Singapore}

\author{Yudai Sato}
\author{Masahiro Haze}
\affiliation{The Institute for Solid State Physics, The University of Tokyo, 5-1-5 Kashiwa-no-ha, Kashiwa 277-8581, Japan}

\author{Junxiang Jia}
\affiliation{Division of Physics and Applied Physics, School of Physical and Mathematical Sciences, Nanyang Technological University, Singapore 637371, Singapore}

\author{K. E. Johnson Goh}
\affiliation{Institute of Materials Research and Engineering (IMRE), Agency for Science, Technology and Research, Singapore}

\author{BaoKai Wang}
\affiliation{Department of Physics, Northeastern University, Boston,
Massachusetts 02115, USA}

\author{Hsin Lin}
\affiliation{Institute of Physics, Academia Sinica, Taipei, Taiwan}

\author{Arun Bansil}
\affiliation{Department of Physics, Northeastern University, Boston,
Massachusetts 02115, USA}

\author{Shantanu Mukherjee}
\affiliation{Department of Physics, Indian Institute Of Technology Madras, Chennai, Tamil Nadu 600036, India}
\affiliation{Computational Materials Science Group
IIT Madras Chennai, Tamil Nadu 600036, India}
\affiliation{Quantum Centres in Diamond and Emergent Materials (QuCenDiem)-Group
IIT Madras Chennai, Tamil Nadu 600036, India}

\author{Yukio Hasegawa}
\affiliation{The Institute for Solid State Physics, The University of Tokyo, 5-1-5 Kashiwa-no-ha, Kashiwa 277-8581, Japan}

\author{Bent Weber}
\email{b.weber@ntu.edu.sg}
\affiliation{Division of Physics and Applied Physics, School of Physical and Mathematical Sciences, Nanyang Technological University, Singapore 637371, Singapore}
\affiliation{ARC Centre of Excellence for Future Low-Energy Electronics Technologies (FLEET), School of Physics, Monash University, Clayton VIC 3800 Australia}

\keywords{Proximity induced superconductivity, quantum spin Hall, Scanning tunneling microscopy, topological insulator}






\newenvironment{sciabstract}{%
\begin{quote} \bf}
{\end{quote}}














\begin{abstract}
The interplay of topology and superconductivity has become a subject of intense research in condensed matter physics for the pursuit of topologically non-trivial forms of superconducting pairing. An intrinsically normal-conducting material can inherit superconductivity via electrical contact to a parent superconductor via the proximity effect, usually understood as Andreev reflection at the interface between the distinct electronic structures of two separate conductors. However, at high interface transparency, strong coupling inevitably leads to changes in the band structure, locally, owing to hybridization of electronic states. Here, we investigate such strongly proximity-coupled heterostructures of monolayer 1T'-WTe$_2$, grown on NbSe$_2$ by van-der-Waals epitaxy. The superconducting local density of states (LDOS), resolved  in scanning tunneling spectroscopy down to 500~mK, reflects a hybrid electronic structure, well-described by a multi-band framework based on the McMillan equations which captures the multi-band superconductivity inherent to the NbSe$_2$ substrate and that induced by proximity in WTe$_2$, self-consistently. Our material-specific tight-binding model captures the hybridized heterostructure quantitatively, and confirms that strong inter-layer hopping gives rise to a semi-metallic density of states in the 2D WTe$_2$ bulk, even for nominally band-insulating crystals. The model further accurately predicts the measured order parameter $\Delta \simeq 0.6$~meV induced in the WTe$_2$ monolayer bulk, stable beyond a 2~T magnetic field. We believe that our detailed multi-band analysis of the hybrid electronic structure provides a useful tool for sensitive spatial mapping of induced order parameters in proximitized atomically thin topological materials.
\end{abstract}

\maketitle

\section{Introduction}

Inducing superconductivity by proximity in materials with non-trivial band topology \cite{FuKane2008, Alicea_2012, beenakker_review_2013} has become a method of choice in the search for unconventional forms of superconducting pairing \cite{mourik_2012signatures, Yazdani2014}. Prominent examples are demonstrations of unconventional superconductivity in semiconductor nanowires with strong spin-orbit coupling \cite{mourik_2012signatures}, atomic chains and islands \cite{Yazdani2014, Palacio2019}, as well as at the surfaces and edges of three-dimensional topological insulators \cite{Sun2016} and semimetals \cite{Yazdani_2019}. At the one-dimensional (1D) edges of 2D topological insulators \cite{FuKane2009, FuKane_QSH_4Pi}, such as the quantum spin Hall (QSH) state \cite{Kane_Mele_2005}, the presence of non-Abelian parafermions have been predicted \cite{Orth2015}. 

\begin{figure*}
\centering
\includegraphics[width=17cm]{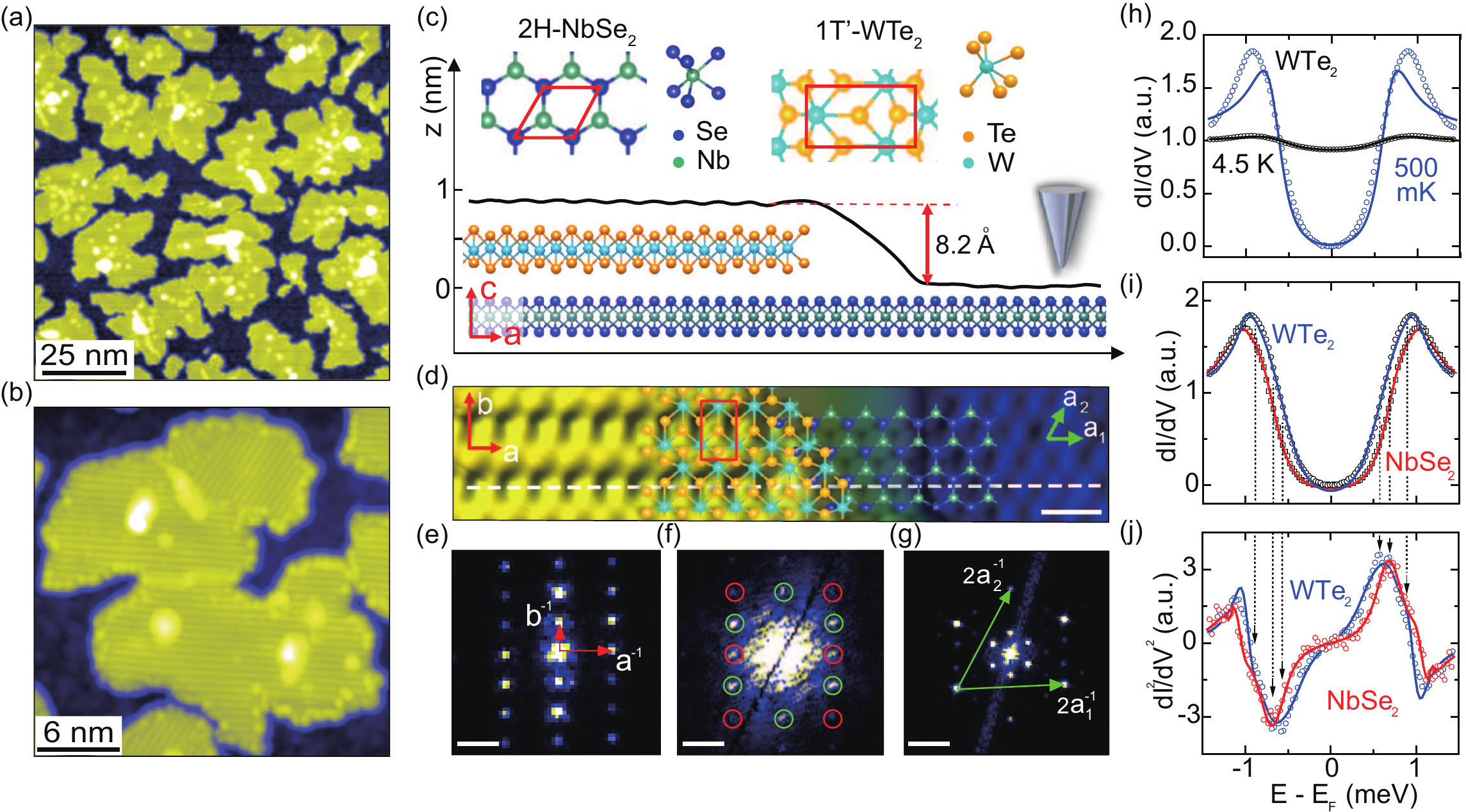}
\caption{(Color online) Superconducting WTe$_2$/NbSe$_2$ heterostructures. (a), (b) STM topography (1.6~V, 100~pA) of monolayer 1T’-WTe$_2$ islands grown on a single crystal of 2H-NbSe$_2$ by van-der-Waals epitaxy. (c) STM measured height profile corresponding to the dashed white line in (d), showing a monolayer height of 8.2~\AA. The insert shows the respective 1T'-WTe$_2$ and 2H-NbSe$_2$ crystal structures. (d) Atomic-resolution close-up (100~mV, 2.2~nA) of the 1T’-WTe$_2$ edge, showing detail of the atomic alignment. The respective chalcogen sublattices are seen to align along the $y$-direction of the 1T’-WTe$_2$ crystal (scale bar: 6.2~\AA). (e)-(g) 2D FFTs of the 1T'-WTe$_2$ lattice (e), an area with 46\% WTe$_2$ monolayer coverage (f), and the pristine NbSe$_2$ surface (g) after cleaving. (all scale bars: 2~nm$^{-1}$). We estimate $a \simeq a_1 = 3.5 \pm 0.2 $~\AA, and $b \simeq 2a_2\textrm{cos}(30\degree) = 6.2 \pm 0.3 $~\AA~ for the lattice constants. (h) Temperature dependence of the superconducting local density of states (LDOS), measured in the WTe$_2$/NbSe$_2$ 2D bulk, comparing measurements at $T = 4.5$~K (black) and 500~mK (blue). Solid blue and black lines are fits to a conventional single-band BCS model. (i) High-resolution tunneling spectra, comparing the WTe$_2$/NbSe$_2$ bulk spectrum with that measured on the bare NbSe$_2$ substrate. Solid red and blue lines are fits to our two-band (NbSe$_2$) and three-band (WTe$_2$) McMillan models, respectively (see main text for detail). (j) Second derivative of the tunnel current, corresponding to the data shown in (i). Dashed lines and arrows indicate the positions of the respective superconducting  energy gaps.}
\label{fig1}
\end{figure*}

In any material system, topologically trivial or non-trivial, the proximity effect can be understood in a microscopic picture to result from Andreev reflection of quasiparticles at the superconductor / normal metal interface. The strength of the induced superconducting pairing is directly linked to the transparency of the interface, i.e. the tunneling coupling at the contact between the proximity-coupled materials. While strong induced pairing with a large order parameter may often be desired, e.g. to mitigate disorder or the effects of an applied magnetic field \cite{mourik_2012signatures}, it is often assumed that the tunnel-coupling does not induce any fundamental changes to the bandstructure. In stark contrast, exchange of charge, especially across atomically abrupt interfaces, can be expected to give rise to atomic-bonding or at least perturbations of the electronic structure, which can lead to profound changes in the electronic band dispersion \cite{Trainer2020}.

Here, we investigate the hybrid electronic structure of such strongly-coupled heterostructures of the quantum spin Hall (QSH) candidate 1T'-WTe$_2$, grown by in-situ van-der-Waals epitaxy on the type-II superconductor 2H-NbSe$_2$. Scanning tunneling spectroscopy down to 500~mK allows us to resolve the superconducting local density of states (LDOS) which we analyze within a self-consistent multi-band framework based on the McMillan equations \cite{mcmillan_1968tunneling}. We are able to show that superconductivity is induced in a semi-metallic WTe$_2$ monolayer bulk, which we understand as a result of a strong inter-layer tunneling across the heterointerface, leading hybridization of electronic states. A material-specific mean-field tight-binding model is able to capture the hybrid electronic structure in the normal state, allowing direct quantitative comparison with the measured LDOS. As a result, our model is able to predict the magnitude of the induced superconducting order parameter, quantitatively, robust to magnetic fields beyond 2~T. Owing to strong hybridization, we observe a significant weakening of the topological edge state signature, measurable by an enhancement of the LDOS in the normal state at the edge, concomitant with a slight enhancement of the induced superconducting order parameter.

Tungsten ditelluride (WTe$_2$) and related members of the transition metal dichalcogenide family with 1T' crystal structure have attracted much attention as candidates material system to realize type-II Weyl \cite{Soluyanov2015} fermions and higher-order topology \cite{Wang2019} in its 3D bulk. Evidence of high-order semi-metallic properties have recently been confirmed in superconducting Josephson junction measurements \cite{Choi2020, Kononov2020} showing that induced supercurrents are well-localized to the materials 1D edges (``hinges''). In atomic monolayers, a time-reversal symmetry protected 2D topologically insulating state \cite{qian_2014quantum} has been predicted, arising from inversion of the transition metal $d$- and the chalcogen $p$-orbitals, highly tunable by electric fields and strain \cite{qian_2014quantum}. While free-standing 1T'-WTe$_2$ monolayers had initially been predicted to be semi-metallic with a negative fundamental band gap \cite{qian_2014quantum}, a positive gap was later confirmed by density functional theory and layer-dependent optical spectroscopy \cite{Zheng2016}. Quantum spin Hall insulating behaviour and the presence of highly-confined 1D metallic states at the edge have since been confirmed in electron transport \cite{fei_2017edge, wu_2018observation}, angle-resolved photoemission spectroscopy (ARPES), and scanning probe measurements \cite{tang_2017quantum, jia2017direct} including microwave impedance microscopy \cite{Shi2019}. However, the sensitivity of the WTe$_2$ band gap to strain \cite{Jia2020-Strain} and electric fields \cite{Maximenko2020} remains a matter of intense debate, as significant variations in the magnitude and definition of the measured QSH bulk are being observed \cite{Zheng2016, fei_2017edge, tang_2017quantum, wu_2018observation, jia2017direct, song2018observation, Maximenko2020}. Recent reports indicate that the bulk gap observed may indeed arise from 2D electronic interactions \cite{song2018observation} challenging the band-insulating picture \cite{Wang2021, Jia2020}. 

The study of superconductivity remains of fundamental interest in both the QSH insulating \cite{qian_2014quantum, Zhang2014, Orth2015} and semi-metallic \cite{Hsu2020, Crepel2021} states of 1T'-WTe$_2$, given theoretical predictions of non-trivial pairing. Intrinsic low-density superconductivity has recently been reported in the 2D bulk of electrostatically doped semi-metallic WTe$_2$ monolayers \cite{Valla_2018gateWTe2, SajadiScience2018}, and is believed to be of non-trivial pairing \cite{Hsu2020, Crepel2021}. Similarly, the 1D helical edge of any quantum spin Hall insulating state has been predicted as a potential host for parafermions in the superconducting state \cite{qian_2014quantum, FuKane_QSH_4Pi, FuKane2009, Zhang2014, Orth2015}. 

Superconductivity up to relatively high critical temperatures and magnetic fields can be achieved by proximity-coupling to a stable intrinsic superconductor, such as the layered type-II superconductor 2H-NbSe$_2$ ($T_C \simeq 7.2$~K, $B_{C2}\simeq 5$~T) \cite{dvir_2018_NbSe2_tunneling, SC_Proxi_Li, SC_proxi_Huang_2018inducing, Hunt_2020proximity}. Further to these studies, here we show that strong coupling to the superconductor gives rise to a hybrid multi-band electronic structure, which needs to be taken into account when interpreting the detailed functional form of the induced superconducting local density of states.

\section{Results and Discussion}

\subsection{Crystal Growth and Atomic Structure}

Figure~\ref{fig1} shows scanning tunneling microscopy (STM) data of 1T'-WTe$_2$/2H-NbSe$_2$ heterostructures, grown by bufferless low-temperature van-der-Waals epitaxy. In agreement with previous studies of WTe$_2$ on bilayer graphene (BLG) \cite{tang_2017quantum, jia2017direct}, we observe Volmer-Weber growth of islands with disordered boundaries (Figs.~\ref{fig1}(a) and \ref{fig1}(b)) and size up to a few tens of nanometre in diameter, that are poly-crystalline on NbSe$_2$ substrates. For this work, we focus on the intermediate coverage limit ($\sim 50$\%) in order to maximize the density of QSH edges, although nearly complete monolayers with coverage of $>$95\% can be achieved with reduced crystal quality. 

Atomic-level detail of the WTe$_2$ lattice alignment with the substrate is shown in Figs.~\ref{fig1}(d)-\ref{fig1}(g) where we find that the respective chalcogen (Te and Se) sublattices align along the atomic rows of 1T'-WTe$_2$ ($y$-direction). A 2D fast Fourier transform (Fig.~\ref{fig1}(f)) of an area with $\sim 50$\% WTe$_2$ monolayer coverage indeed confirms that the Bragg peaks of the WTe$_2$ and NbSe$_2$ lattices coincide within the accuracy of the measurement. We extract $a \simeq a_1 = 3.5 \pm 0.2 $~\AA, and $b \simeq 2a_2\textrm{cos}(30\degree) = 6.2 \pm 0.3$~\AA~, in good agreement with the lattice parameters of WTe$_2$ ($a = 3.48$~\AA, and $b = 6.28$~\AA~) \cite{Zheng2016}. $a_1 = a_2 = 3.45$~\AA are the lattice parameters of NbSe$_2$, local lattice matching would imply a 5\% compressive lattice strain along $b$, which has previously been shown to further stabilize the WTe$_2$ bulk gap \cite{Jia2020-Strain}.

A first indication that we may expect strong hybridization in WTe$_2$/NbSe$_2$ heterostructures comes from a measurement of the monolayer height $h = 0.9 \pm 0.1$~nm, extracted from $z$-height distributions of large-scale topographic STM images (see Fig.~\ref{histogram}, Appendix A), and is comparable to measurements of van-der-Waals stacked WTe$_2$/NbSe$_2$ ($h \simeq 0.7$~nm) \cite{Hunt_2020proximity}. The significantly lower layer height, compared to WTe$_2$/BLG ($h \simeq 1.2$~nm) \cite{jia2017direct} and WTe$_2$/HOPG ($h = 1.2 \pm 0.1$~nm) (this work), suggests a smaller van-der-Waals gap, resulting in stronger interlayer coupling and hybridization of electronic states.

We observe some variations in the electronic structure across different monolayer crystals, reflecting varying doping levels and the presence of in-gap impurity states around the visible adatom disorder (Figs.~\ref{fig1}(a)-(b). We therefore focus on spectra obtained on clean monolayer regions which display a clear suppression of the LDOS over an energy range of several tens of meV around the Fermi energy, and with signatures of an edge state in the normal state. As shown in the comparison of Fig.~\ref{fig2} below, such spectra agree well with tight-binding calculations of the normal state electronic structure, as well as with those published on WTe$_2$/BLG \cite{tang_2017quantum, jia2017direct, Jia2020-Strain, Maximenko2020}. Given the metallic nature of the two substrates used in this work, we do not expect that exitonic insulating effects \cite{Wang2021, Jia2020} play a dominant role here, as Coulomb interaction would be expected to be strongly screened.

\subsection{Multi-Band Superconductivity}\label{sec:B}

\begin{figure*}[htb]
\centering
\includegraphics[width=18cm]{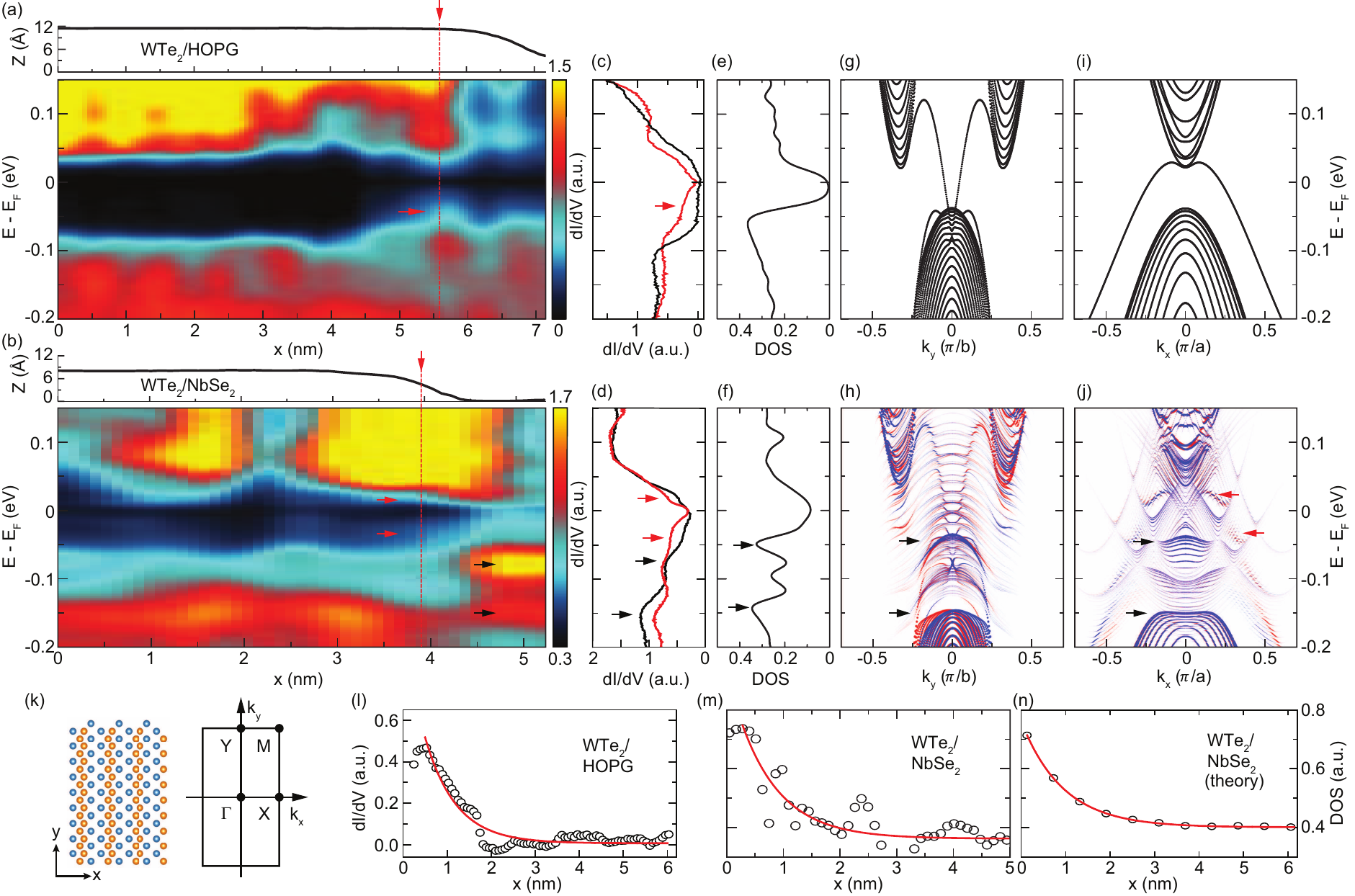}
\caption{(Color online) {Hybrid electronic structure of strongly-coupled 1T'-WTe$_2$ heterostructures.} (a)-(b) Spatial profiles of the measured normal-state local density of states (LDOS) across the monolayer edge for (a) WTe$_2$/HOPG and (b) WTe$_2$/NbSe$_2$, alongside corresponding STM height profiles. (c),(d) individual point spectra of the local density of states (LDOS) in 2D bulk (black) and edge (red), measured at $T=4.2$~K, and compared to tight-binding calculation (e, f), respectively. (g)-(j) Tight-binding band structures of free-standing WTe$_2$ monolayers in $\Gamma - Y$ (g) and $\Gamma - X$ (i) direction, respectively, and compared to the spin-resolved (blue/red) WTe$_2$ orbital weight of a WTe$_2$/NbSe$_2$ heterostructure in $\Gamma - Y$ (h) and $\Gamma - X$ (i). The measured LDOS in (a) and (b) was taken along $y$-direction, across the $\Gamma - X$ edge. Red dashed lines in (a) and (b) indicate the positions at which the edge state point spectra in (c) and (d) were measured. Black arrows indicate spectral features resolved on WTe$_2$/NbSe$_2$, only, arising from the NbSe$_2$ substrate. Red arrows indicate the enhanced LDOS at the crystal edges. (k) Lattice model and first Brillouin zone of 1T'-WTe$_2$, indicating high-symmetry points and directions. (l)-(n) Spatial profiles of the integrated edge-state LDOS (-50 meV to +20 meV) for (l) WTe$_2$/HOPG and (m) WTe$_2$/NbSe$_2$, compared to tight-binding calculations (n). We define the edge position, electronically, where the integrated LDOS peaks and exponentially decays into the 2D bulk gap. Red lines in (l)-(n) are exponential fits to extract the edge state decay lengths, $\xi = (0.7 \pm 0.1)$~nm (WTe$_2$/HOPG), $\xi = (0.7 \pm 0.2)$~nm (WTe$_2$/NbSe$_2$), and $\xi = (1.0 \pm 0.0)$~nm (theory).}
\label{fig2}
\end{figure*}

The clear signature of a superconducting energy gap, measured in the 2D bulk of monolayer WTe$_2$, is shown in Figs.~\ref{fig1}(h), comparing the superconducting LDOS at 4.5~K with that measured at 500~mK. Fits to conventional BCS theory (solid lines) describe the data well only at 4.5~K, but fail to describe the details of the energy gap resolved at 500~mK. This becomes particularly apparent by an underestimation of the width and height of the coherence peaks at $\pm 1$~meV. A comparison of high-resolution spectra measured, respectively, on the WTe$_2$ and NbSe$_2$ surfaces, are shown in Fig.~\ref{fig1}(i). We observe a slightly reduced gap size on NbSe$_2$ post MBE growth, compared to the pristine NbSe$_2$ surface, which is likely due to a partial quenching of the 3Q charge density wave (CDW) order \cite{noat_2010signatures, noat_2015_2band} leading to disorder averaging of the Fermi surface anisotropy and reduced inter-band coupling (see Fig.~\ref{CDW} and Appendix B for further detail). A further reduction of the superconducting gap measured on WTe$_2$/NbSe$_2$ corroborates that superconductivity is induced into the monolayer via the substrate.

As detailed below, we are able to accurately describe the measured superconducting LDOS within a self-consistent multi-band framework based on the McMillan equations \cite{mcmillan_1968tunneling, Japanese_2004_MgB2, noat_2015_2band} (blue/red lines). In the presence of multiple \textit{non-interacting} bands $i$, the superconducting density of states can be expressed as a simple sum $N(E) = \sum\limits_{i} N_i(E)$ of two or more partial densities of states,
\begin{equation}
    N_i(E) = \tilde{N_i}(E_F) \operatorname{Re} \left[\frac{\lvert{E - i\gamma_i}\rvert}{\sqrt{(E - i\gamma_i)^2 - \Delta_i(E)^2}}\right]
\label{eq:Ni}
\end{equation}
Within conventional BCS theory, the order parameter $\Delta$ is a constant and the prefactor. The prefactor, $\tilde{N_i}(E_F)$), denotes the partial state density of states of band $i$ in the normal state and, in the superconducting state, takes the role \cite{noat_2015_2band} of an effective partial density of states weight to the total density of states (Eq.~(\ref{totalDOS})). An additional broadening parameter $\gamma_i$ is usually included to account for energy broadening during the tunneling process (Dynes parameter) \cite{dynes_1978_function_direct}.

Different from BCS theory, the McMillan model \cite{mcmillan_1968tunneling} considers \textit{coupled} bands with order parameters $\Delta_i(E)$, that are energy-dependent complex functions of the form 
\begin{dmath}
\Delta_i(E) = \frac{\Delta_{i}^0 + \Gamma_{ij}\Delta_{j}(E)/\sqrt{\Delta_{j}^2(E)-(E -i\gamma_{j})^2}}{1 + \Gamma_{ij}/\sqrt{\Delta_{j}^2(E) - (E -i\gamma_{j})^2}}
\label{eq3}
\end{dmath}
Here, an inter-band Cooper pair tunneling across the van-der-Waals gap at rate $\Gamma_{ij}$ renormalizes the intrinsic order parameters of the separate materials $\Delta_{i}^0$, leading to a set of self-consistent equations, that can be solved numerically to fit our data. 

Although originally developed to describe the superconducting proximity effect in single-band metallic systems, the McMillan model has more recently been successfully applied to describe intrinsic multi-band superconductivity in layered two-band superconductors \cite{Japanese_2004_MgB2}, including NbSe$_2$ \cite{noat_2015_2band, dvir_2018spectroscopy}. For this work, we further extend the McMillan model to describe the hybrid electronic structure of 1T'-WTe$_2$/NbSe$_2$ heterostructures, thus treating multi-band superconductivity and proximity-coupling within the same theoretical framework. To this end, we consider a third order parameter $\Delta_{\rm WTe_2}$, coupled to the NbSe$_2$ $K$-band (see supporting information), assuming $\Delta_{\rm WTe_2}^0=0$. Fits to both a two-band (NbSe$_2$, red line) and a three-band (WTe$_2$/NbSe$_2$, blue line) model describe the data well (Fig.~\ref{fig1}(i)), and reproduce the known NbSe$_2$ order parameters (see Table~\ref{table-NbSe2}. Indeed, we find that simpler BCS-based models are unable to represent the data well at 500~mK (Appendix B), further confirming the multi-band hybrid electronic structure. For the real part of the induced order parameter, we find $\Delta_{\rm WTe_2}(E_F) = (0.57 \pm 0.02)$~meV, reflecting induced pairing in the strongly coupled heterostructure, in reasonable agreement with recent transport \cite{SC_proxi_Huang_2018inducing} and local probe \cite{Hunt_2020proximity} spectroscopy of non-epitaxial WTe$_2$/NbSe$_2$ hetero-junctions.

\subsection{Hybrid Electronic Structure}\label{hybrid}

As inferred from our three-band model, we understand that superconductivity is induced in the WTe$_2$ bulk as a result of strong inter-band coupling to the substrate giving rise to a WTe$_2$/NbSe$_2$ hybrid electronic structure. We confirm this notion in Fig.~\ref{fig2} by directly comparing spectra of the normal-state LDOS with tight-binding band structure calculations \cite{lau19} (Appendix F). In Figs.~\ref{fig2}(a) and \ref{fig2}(b) we show the spatial evolution of the measured normal-state LDOS across a monolayer edge, comparing the WTe$_2$/HOPG (a) and the WTe$_2$/NbSe$_2$ (b) heterostructure, alongside corresponding height-profiles. Individual point spectra of the 2D bulk (black) and 1D edge (red) are furthermore shown in the inserts to Figs.~\ref{fig2}(c) and (d). In both heterostructures, the LDOS shows clear signs of a suppression over $\simeq 70$~meV around the Fermi energy, corresponding to a soft gap. This gap is significantly more developed in WTe$_2$/HOPG, but shows with comparable magnitude and position also on WTe$_2$/NbSe$_2$. A residual 2D bulk LDOS within the gap is measurable on both substrates, but is much more pronounced in WTe$_2$/NbSe$_2$, attesting to strong hybridization in the heterostructure.

Tight-binding band structure calculations of a freestanding WTe$_2$ monolayer are shown in Fig.~\ref{fig2}(c), and compared with the spin-resolved WTe$_2$ orbital weight of the hybridized WTe$_2$/NbSe$_2$ heterostructure (Fig.~\ref{fig2}(d). We find best agreement for an interlayer hopping of 0.15~eV (see Appendix), similar in magnitude to the Nb-Nb hopping strength within the 2H-NbSe$_2$ cell \cite{Rahn2012}. A slightly larger hopping strength would be expected given the shorter Nb-Te separation. The direct comparison with our measured spectra (inserts) shows reasonable agreement with regard to the position of the band edges (horizontal dashed lines), the Fermi energy ($E = E_F$), as well as edge state features (red arrows). Indeed, the WTe$_2$/HOPG spectra resemble closely those previously reported for monolayer WTe$_2$ grown on bilayer graphene, \cite{tang_2017quantum, jia2017direct} indicating that heterostructures with graphitic substrates are only weakly hybridized. The additional spectral features (black arrows), however, only observed in the WTe$_2$/NbSe$_2$ heterostructure, need to be attributed the presence of a NbSe$_2$ partial DOS (reference spectrum in the left-hand insert to Fig.~\ref{fig2}(d)), confirming the hybridization picture. 

As a result of the strong hybridization in WTe$_2$/NbSe$_2$, we further observe a substantial weakening of the topological edge state signature, compared to our observations on WTe$_2$/HOPG and WTe$_2$/BLG\cite{tang_2017quantum, jia2017direct, Jia2020-Strain, Maximenko2020}. This is illustrated in Figs.~\ref{fig2}(e)-\ref{fig2}(g) by plotting spatial profiles of the integrated LDOS within the gap (-50 meV and +20 meV) away from the edge position. We extract roughly consistent exponential decay lengths of $\sim 1$~nm for both heterostructures. However, the ratio of the edge state LDOS to that in the bulk is much lower for WTe$_2$/NbSe$_2$ ($\simeq 1.8$), compared to WTe$_2$/HOPG ($\simeq 10$), given the large residual LDOS in the bulk gap. Both edge state decay length and LDOS ratio agree well with our tight-binding calculations of the hybrid electronic structure for WTe$_2$/NbSe$_2$ as shown in Fig.~\ref{fig2}(g).

\begin{figure}[htb]
\centering
\includegraphics[width=\columnwidth]{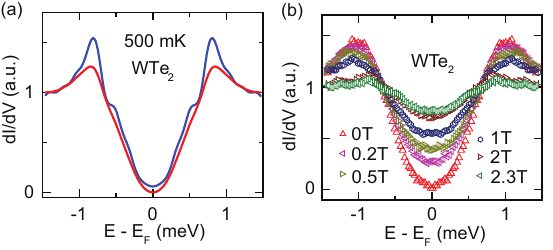}
\caption{(Color online) Magnetic field dependence of the WTe$_2$/NbSe$_2$ superconducting local density of states (LDOS). (a) Calculated LDOS at 500~mK for an interlayer coupling of 0.15~eV with (red) and without (blue) energy broadening (0.15~meV) included. (b) Magnetic field dependence of the measured superconducting LDOS up to $B=2.3$~T, with magnetic field applied perpendicular to the heterostructure ($c$-axis).}
\label{fig3}
\end{figure}

In Fig.~\ref{fig3}, we plot the calculated superconducting LDOS for the same inter-layer hopping (0.15~eV) as used for both the band structure calculation in Fig.~\ref{fig2}(d) and the LDOS profile in Fig.~\ref{fig2}(g). We extract $\Delta = 0.58$~meV on the top-most tellurium sublattice, which the STM is expected to be most sensitive to, in remarkable agreement with the experiment ($\Delta_{\rm WTe_2}(E_F) = (0.57 \pm 0.02)$~meV), as extracted from our three-band fits. 

The magnetic field stability of superconducting pairing in the WTe$_2$ monolayer bulk is investigated in Fig.~\ref{fig3}(b), where we plot the measured superconducting LDOS for magnetic fields up to $B= 2.3$~T, applied perpendicular to the sample plane ($c$-axis). 

\subsection{Spatial Profile at the Monolayer Edge}

\begin{figure}[htb]
\centering
\includegraphics[clip,width=\columnwidth]{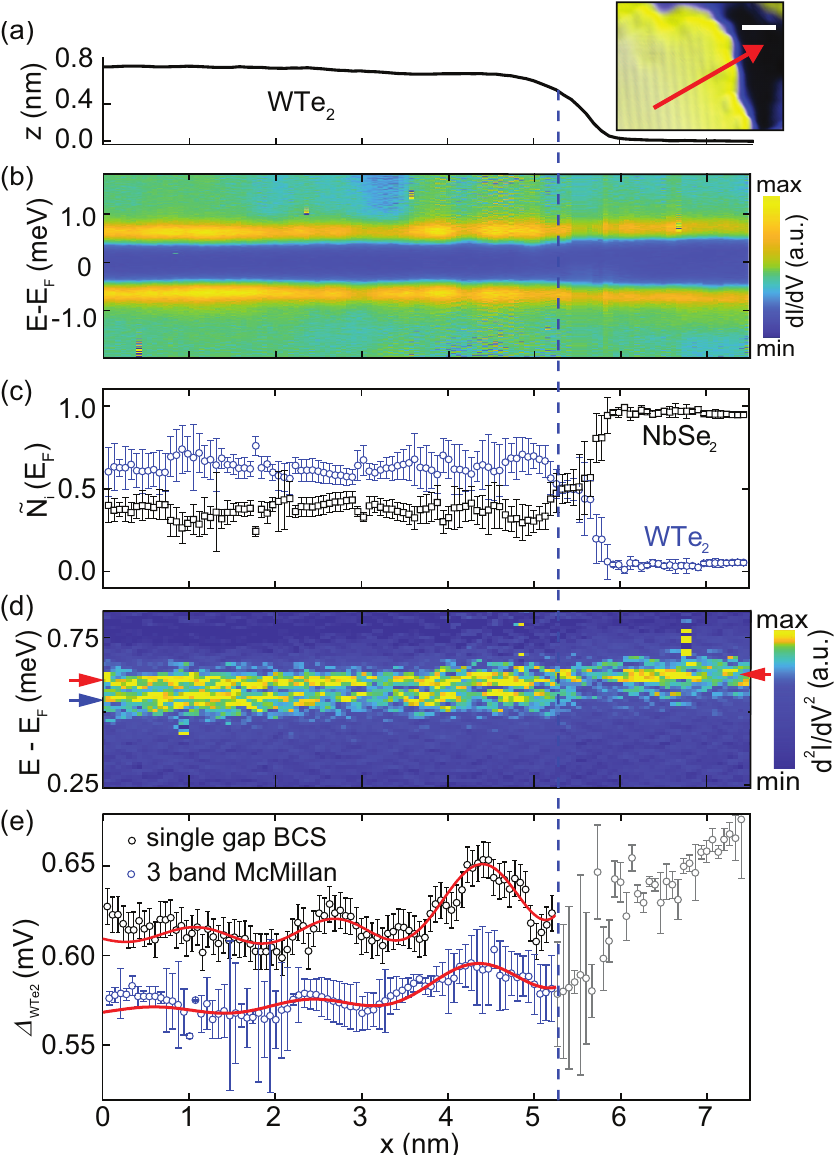}
\caption{(Color online) Multi-band superconductivity at the WTe$_2$ monolayer edge. (a) STM-height profile measured across a clean edge of a WTe$_2$ monolayer crystal. The insert shows the corresponding STM topograph indicating position and direction of the line profile (red arrow). (b) A series of tunneling spectra measured $T = 500$~mK for points across the edge. (c) The partial density of states weight $\tilde{N_i}(E_F)$, extracted from our three-band analysis of (b), shows a clear crossover at the edge $x_0 \simeq 5.3$~nm (dashed line) beyond which the WTe$_2$ partial density of states vanishes. (d) The second derivative of the tunnel current close to the coherence peaks shows clear signatures of three-band superconductivity. (e) Extracted spatial profile of the induced order parameter $\Delta_{\rm WTe_2}(x)$ (blue), compared to a single-band BCS/Dynes model (black). Solid red lines are fits to Eq.~(3).}
\label{fig4}
\end{figure}

In Fig.~\ref{fig4}, we investigate the spatial evolution of the superconducting LDOS across the WTe$_2$ monolayer edge. An atomic resolution STM image of a clean WTe$_2$ edge is shown in Fig.~\ref{fig4}(a), indicating position and direction of the STM height profile. The spatial dependence of the superconducting energy gap, measured at 500~mK, is shown Fig.~\ref{fig4}(b), alongside the superconducting partial density of states weight $\tilde{N_i}(E, E_F)$ of the WTe$_2$/NbSe$_2$ heterostructure (Fig.~\ref{fig4}(c)). The latter has been extracted self-consistently, using Eqns.~(\ref{eq:Ni}) and (\ref{eq:delta}). We observe a pronounced crossover in the partial density of states at ($x_0 \simeq 5.3$~nm), coinciding with the edge position, beyond which the WTe$_2$ DOS drops to zero as the STM tip leaves the monolayer crystal. In the monolayer bulk, we find that WTe$_2$ dominates the total DOS with $(63 \pm 5)$\%, with a contribution of $(37 \pm 5)$\% by the NbSe$_2$ substrate partial DOS. Interestingly, a more pronounced enhancement in the superconducting partial DOS at the edge appears absent, at least when compared to the normal-state DOS in Fig.~\ref{fig2}(f). This is likely due to the vanishing total LDOS at the edge at $E-E_F \simeq 0$ seen in the normal state (compare red curves in Figs.~\ref{fig2}(c) and \ref{fig2}(d), possibly due to the presence of an interaction driven pseudogap \cite{stuhler2020, reis_2017bismuthene, tang_2017quantum, fuhrer_2018_Na3Bi}. 

A further confirmation of three-band superconductivity is given in Fig.~\ref{fig4}(d) where we plot the spatial profile of the second derivative of the tunneling current for energies close to the coherence peak. We observe a sharp transition from two-band to three-band signatures as soon as the probe tip crosses the monolayer edge (dashed line). Arrows indicate the gap energies for the WTe$_2$ induced gap (blue) and the intrinsic NbSe$_2$ small gap (red). The NbSe$_2$ large gap at higher energy is not seen here, but is indicated in Figs.~\ref{fig:models}(a) and \ref{fig:models}(b) of the Appendix. 

A spatial profile of the extracted order parameter $\Delta_{\rm WTe_2}(x)$ is shown in Fig.~\ref{fig4}(e), in which we observe spatial oscillations with a period of $\simeq 2$~nm. Oscillations of comparable period are also observed in the order parameter extracted from a simple single-band BCS model, as well as are visible in the energy-integrated normal-state LDOS (Fig.~\ref{fig2}(f)). This suggests that these arise from Friedel-like oscillations in the local density of states due to scattering of 2D bulk states at the WTe$_2$ edge. Indeed, we can fit the extracted order parameter with a simple empirical model,
\begin{dmath}
\Delta(x) = \frac{A\sin{[2k_F(x-x_0)]}}{(x-x_0)} + B\exp\left(\frac{x-x_0}{\xi}\right) + C
\end{dmath}\label{Friedel}
in which the first term reflects the oscillatory behaviour in the local order parameter due to quasiparticle-interference in the 2D bulk. The second term accounts for a residual enhancement of the exponentially decaying 1D edge state, with decay length $\xi = (1.1 \pm 0.2)$~nm comparable to our measurements of the normal state LDOS (Fig.~\ref{fig2}(g)). The simple model thus simultaneously confirms the edge position ($x_0 = 5.3$~nm) and the Fermi wave vector\cite{Wang2017velocity} $k_{\rm F} \simeq 1.8$~nm$^{-1}$, and the edge state decay length. Upon crossing the edge, $\Delta_{\rm WTe_2}(E_F)$ increases to approach the smaller of the two renormalized order parameters of NbSe$_2$ ($\Delta_S(E_F)= 0.68 \pm 0.08$~meV) as the three-band model emulates the two-band superconductivity of the substrate. This transition coincides with a vanishing WTe$_2$ partial DOS, which confirms that the greyed-out data points in Fig.~\ref{fig4}(e) do not contribute any significant spectral weight to the overall measured density of states.

\section{Conclusions}

In summary, we have reported signatures of multi-band superconductivity in strongly-coupled 1T'-WTe$_2$/NbSe$_2$ heterostructures, grown by van-der-Waals epitaxy. Analyzing the superconducting density of states down to 500~mK in scanning probe spectroscopy, we have shown that strong hybridization of electronic states gives rise to a semimetallic density of states in the 2D bulk even in nominally band-insulating crystals. Describing the detailed functional form of the superconducting energy gap in a self-consistent multi-band framework, based on the McMillan equations, we confirm the strong inter-band coupling. Our quantitative comparison of the measured local density of states with a material-specific tight-binding model ultimately confirms the hybrid electronic structure, in both normal and superconducting states for the same interlayer hopping, thus accurately predicting the magnitude of the induced WTe$_2$ order parameter, $\Delta_{\rm WTe_2} (B=0) \simeq 0.55$~meV, stable beyond a 2~T magnetic field. Despite the strong hybridization, we find that a measurable enhancement of the measured local density of states persists at the crystal edges, detectable in both the normal state and in a slight enhancement of the  order parameter in the superconducting state. We believe that our multi-band treatment of strongly hybridized van-der-Waals heterostructures will form a useful tool to mapping spatial variation of the induced superconducting order parameter in wider range proximitized atomically-thin topological materials \cite{Lodge2021}.

\section*{Acknowledgements}

This research is supported by National Research Foundation (NRF) Singapore, under the Competitive Research Programme "Towards On-Chip Topological Quantum Devices" (NRF-CRP21-2018-0001), with partial support from a Singapore Ministry of Education (MOE) Academic Research Fund Tier 3 grant (MOE2018-T3-1-002). The work was supported in part by Grants-in-Aid for Scientific Research from the Japan Society for the Promotion of Science (nos. 16H02109, 18K19013, and 19H00859). The work at Northeastern University was supported by the US Department of Energy (DOE), Office of Science, Basic Energy Sciences grant number DE-SC0019275 and benefited from Northeastern University's Advanced Scientific Computation Center (ASCC) and the NERSC supercomputing center through DOE grant number DE-AC02-05CH11231. KEJG acknowledges support from the Agency for Science, Technology and Research (A*STAR) under its A*STAR QTE Grant No. A1685b0005. H.L. acknowledges the support by the Ministry of Science and Technology (MOST) in Taiwan under grant number MOST 109-2112-M-001-014-MY3. SM would like to acknowledge the new faculty seed grant from IIT Madras under project number Project No: PHY/18-19/703/NFSC/SHAA. BW acknowledges a Singapore National Research Foundation (NRF) Fellowship (NRF-NRFF2017-11). We thank Jack T. Hellerstedt for technical advice during the early stages of crystal growth.
\medskip

\section*{Appendix A: 1T'-WTe$_2$ Crystal Growth}

\begin{figure}
\centering
\includegraphics[clip,width=\columnwidth]{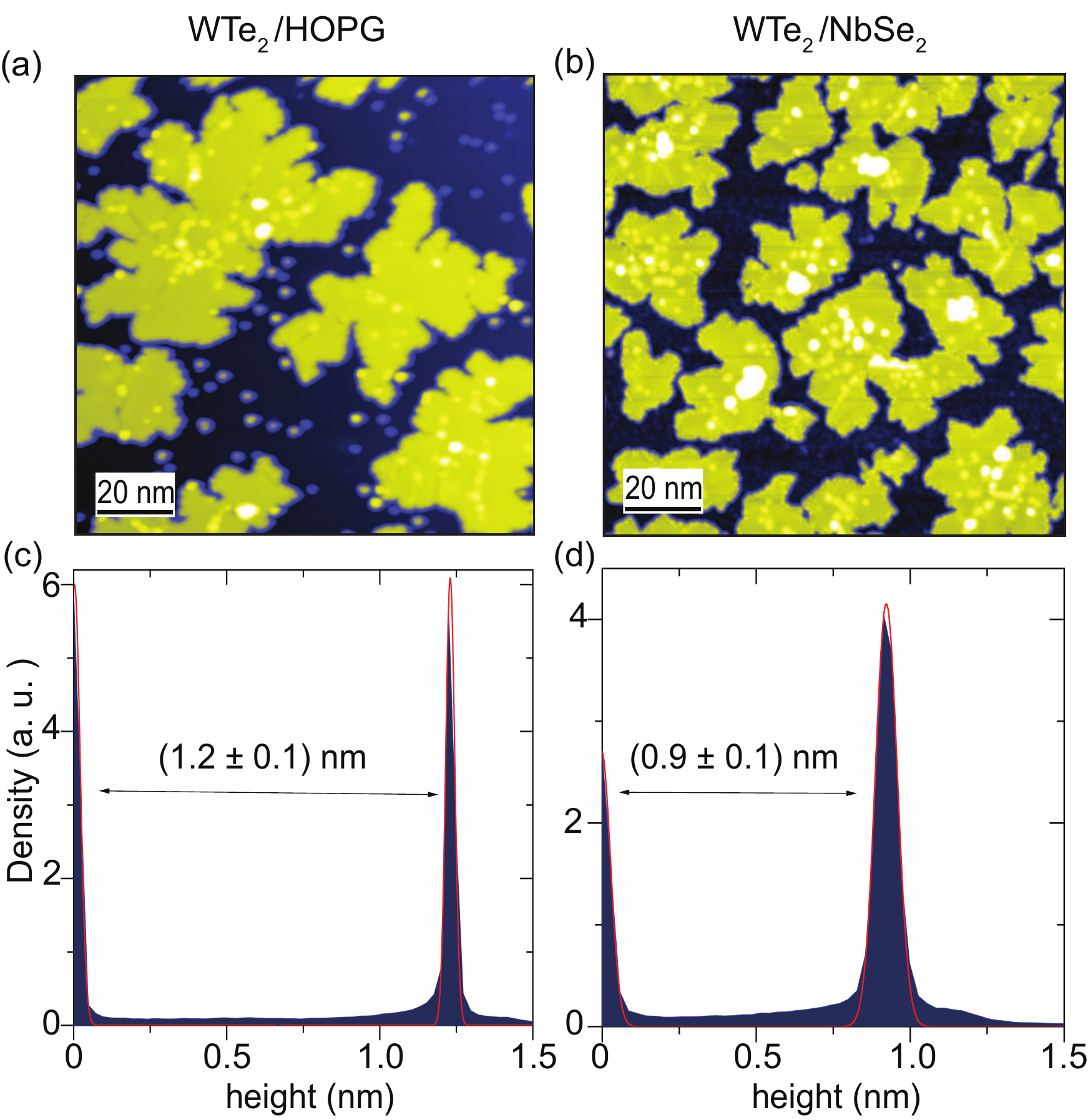}
\caption{(Color online) Height estimates for monolayer WTe$_2$ islands on different substrates. (a)-(b) Large-scale STM topography images, showing a number of 1T'-WTe$_2$ islands, respectively, on HOPG (a) and on NbSe$_2$ (b) substrates. (c)-(d) Corresponding histograms of the $z$-height information extracted. Fits to a normal distribution (red lines) allows to estimate the monolayer height $(1.2 \pm 0.1)$~nm (WTe$_2$/HOPG) and $(0.9 \pm 0.1)$~nm (WTe$_2$/NbSe$_2$), respectively.
}
\label{histogram}
\end{figure}

Van-der-Waals molecular beam epitaxy (MBE) was performed in an Omicron Lab10 MBE (base pressure $2 \times 10^{-10}$~mBar) on bulk crystals of 2H-NbSe$_2$ (HQ graphene, Netherlands) and highly oriented pyrolytic graphite (HOPG). The substrates were mechanically cleaved in UHV after degassing overnight (300\degree~C), following co-deposition of W (99.998$\%$) and Te (99.999$\%$) for approximately 1h at a Te:W flux ratio of roughly 260:1. We use slightly different substrate temperatures for the two heterostructures with 145\degree~C for 1T’-WTe$_2$/NbSe$_2$, and 230\degree~C for 1T’-WTe$_2$/HOPG. Typical STM images of the WTe$_2$/HOPG and WTe$_2$/NbSe$_2$ heterostructures, respectively, are shown in Fig.~\ref{histogram}(a) and \ref{histogram}(b), alongisde measurements of the respective WTe$_2$ monolayer height, extracted from $z$-height histograms \ref{histogram}(c) and \ref{histogram}(d).

\section*{Appendix B: Electronic Structure and Superconductivity of NbSe$_2$}

NbSe$_2$ is a type-II superconductor \cite{Wilson1977, Wexler_1976} with a critical temperature $T_C = 7.2$~K and field $B_{\rm C2}\sim 5$~T \cite{ NbSe2_critical_temp, Foner1973-NbSe2-field}. The microscopic detail of its superconductivity have been investigated, intensively, both theoretically and experimentally \cite{noat_2010signatures, noat_2015_2band, dvir_2018_NbSe2_tunneling, khestanova_2018unusual}.

\begin{table}
\caption{\textbf{Multi-band superconductivity and Fermi-surface anisotropy.} Summary of measured NbSe$_2$ order parameters published in literature.}
\begin{center}
\begin{tabular}{|c||c|c|c|}
\hline
Model & \thead{Measured Gap \\ (meV)} & Reference & Experiment\\
 \hline
   & \thead{$\Delta_S=0.50$ \\ $\Delta_L=1.05$} & [\onlinecite{Hunt_2020proximity}] & STS \\
 \cline{2-4}
  \thead{Two-band \\ (BCS)} & \thead{$\Delta_S=0.73$ \\ $\Delta_L=1.26$} & [\onlinecite{SC_proxi_Huang_2018inducing}] & \thead{Specific \\ Heat} \\
 \cline{2-4}
  &  \thead{$\Delta_S=0.85$ \\ $\Delta_L=1.50$} & [\onlinecite{Haihu2008}] & \thead{Specific \\ Heat}\\
 \cline{1-4}
  & \thead{$\Delta_{\rm min} = 0.65$ \\ $\Delta_{\rm max} = 1.62$} & [\onlinecite{Haihu2008}] & \thead{Specific \\ Heat}\\  
 \cline{2-4}
 \thead{Single-band \\ (anisotropic)} & \thead{$\Delta_{\rm min} = 0.56$ \\ $\Delta_{\rm max} = 1.30$} & [\onlinecite{fletcher2007}] & \thead{Penetration \\ length} \\
 \cline{2-4}
  & \thead{$\Delta_{\rm min} = 0.62$ \\ $\Delta_{\rm max} = 1.30$} & [\onlinecite{khestanova_2018unusual}] & \thead{Electron \\ transport}  \\
 \hline
  & \thead{$\Delta_S=0$ \\ $\Delta_L=1.3$} & [\onlinecite{noat_2015_2band}] & STS \\
 \cline{2-4}
  \thead{Two-band \\ McMillan} & \thead{$\Delta_S=0 - 0.3$ \\ $\Delta_L=1.26$} & [\onlinecite{dvir_2018spectroscopy}] & \thead{Electron \\ transport}\\
\cline{2-4}
    & \thead{$\Delta_S =0.45$ \\ $\Delta_L = 1.10$} & this work & STS\\ \hline
\end{tabular}
\end{center}\label{table-NbSe2}
\end{table}

\begin{figure}[htb]
\centering
\includegraphics[width=7.0cm]{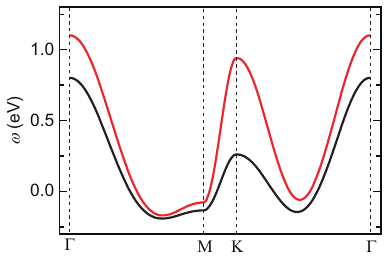}
\caption{(Color online) Tight-binding calculation of the low-energy NbSe$_2$ band structure. The black and red lines show bands arising from the $d_{3z^2-r^2}$ orbitals, centered on the two Nb atoms within the unit cell.}
\label{fig:NbSe2}
\end{figure}

The NbSe$_2$ Fermi surface \cite{Borisenko2009,Rahn2012,Yokoya2001} is composed of at least two bands at the Fermi energy. Tight-binding calculations of the NbSe2$_2$ low-energy bandstructure are shown in Fig.~\ref{fig:NbSe2}, where black and red lines are bands due to the $d_{3z^2-r^2}$ orbitals, centered on the two Nb atoms within the unit cell (Nb bands). These form bonding and anti-bonding orbitals, giving rise to cylinders at the six $K$-points of the hexagonal Brillouin zone. A third band (predominantly Se) \cite{noat_2010signatures} has been neglected in our model, as it is not believed to contribute to multi-band superconductivity in NbSe$_2$ \cite{noat_2010signatures}. As a result of the two-band nature, two distinct superconducting energy gaps, $\Delta_L$ and $\Delta_S$, are usually observed at very low temperatures ($T \ll T_C$) \cite{noat_2015_2band,SC_proxi_Huang_2018inducing, Haihu2008} (see Table~\ref{table-NbSe2}), each hosted within the two Nb bands \cite{noat_2010signatures, noat_2015_2band, dvir_2018spectroscopy}. Superconducting pairing is significantly stronger in the Nb $K$-bands \cite{noat_2010signatures, noat_2015_2band, dvir_2018_NbSe2_tunneling}, giving rise to the larger of the two gaps $\Delta_L \simeq 1.3$~meV. At $\Gamma$, the pairing is weaker with order parameter $\Delta_S$, and is usually found ranging between $0 - 0.3$~meV \cite{noat_2015_2band, dvir_2018_NbSe2_tunneling}. 

\begin{figure}
\centering
\includegraphics[clip,width=8cm]{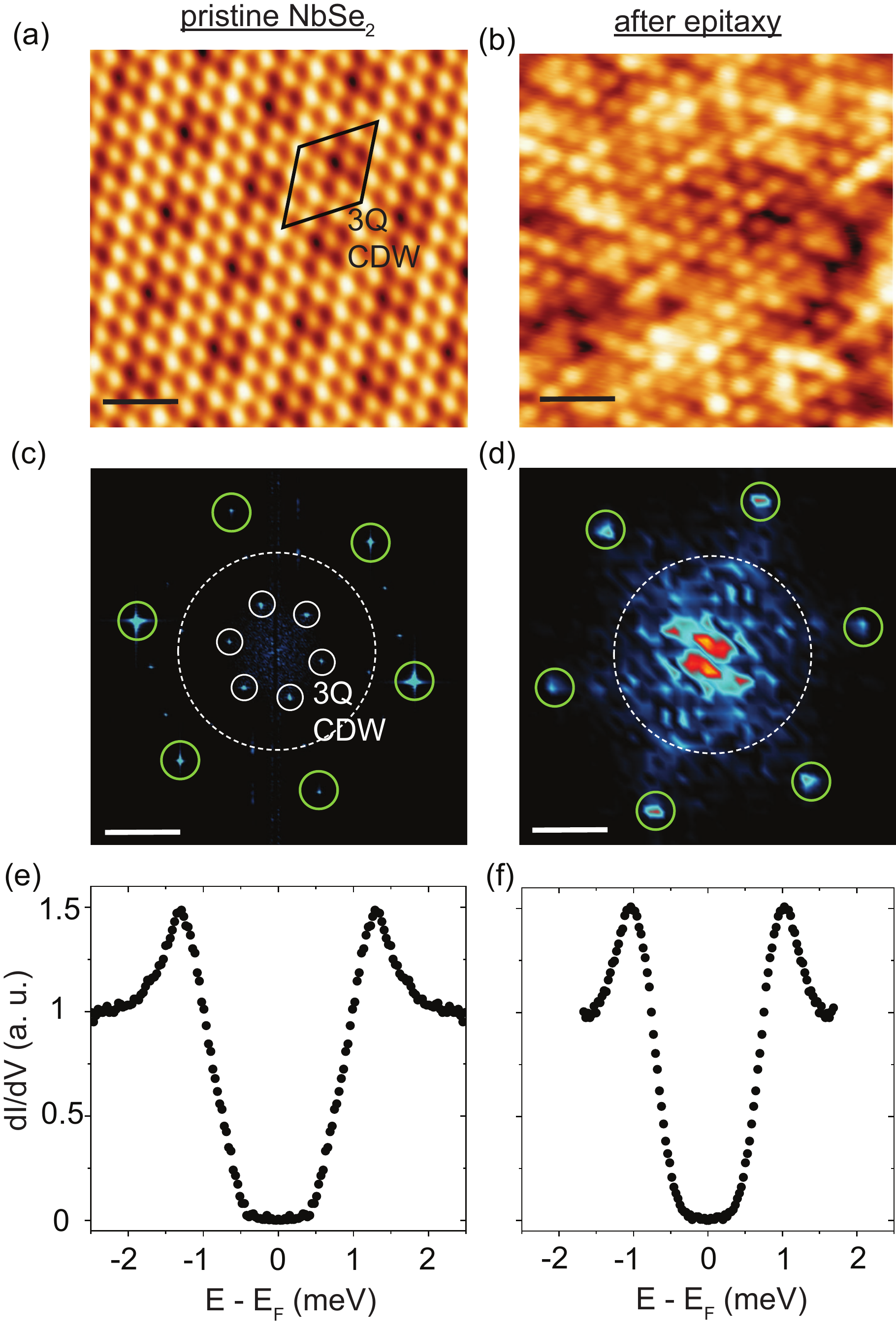}
\caption{(Color online) Quenching of charge density wave (CDW) order on the NbSe$_2$ surface post-epitaxy. (a,b) STM images of the pristine NbSe$_2$ surface after cleaving in UHV (a), and following epitaxy of a sub-monolayer coverage of WTe$_2$ (b). (c,d) Corresponding FFTs show that clear signatures of the 3Q CDW wave vector (c) cannot be discerned post-epitaxy (d). (e,f) Corresponding measured local density of states, measured before (e) and after (f) epitaxy, showing a slight reduction and rounding of the superconducting energy gap.}
\label{CDW}
\end{figure}

CDW order on the pristine NbSe$_2$ surface also plays an important role in scanning tunneling microscopy measurements of the superconducting gap. In pristine NbSe$_2$, the 3Q charge density wave (CDW) gives rise to an anisotropic Fermi surface and a reduced Brillouin zone with hexagonal symmetry \cite{noat_2015_2band}. Possible mechanism for a suppression of $T_C$ have been argued to result from a reduction of the highly anisotropic superconducting gap due to disorder averaging and/or from the suppression of assistive short range CDW correlations \cite{cho2019}. From a multi-band perspective, the CDW wave vectors $g_{i,\rm CDW} = 1/3 G_i$ link states at the $\Gamma$ point (associated with the NbSe$_2$ small gap) to the Nb cylinders at $K$ (associated with the large gap). This allows for the observation of the large gap on pristine NbSe$_2$ surfaces in scanning tunnelling experiments \cite{noat_2010signatures, noat_2015_2band}, which are most sensitive to states at $\Gamma$ (tunnelling of electrons with finite transverse momentum is strongly suppressed in STM due to the vertical tunneling path). Once CDW order is suppressed, STM will predominantly probe states at $\Gamma$, which would explain the reduced gap observed \cite{noat_2015_2band}. Figure \ref{CDW} shows a comparison of the measured gap on both the pristine and the post-growth NbSe$_2$ surface, with a reduction in gap size clearly visible.

\section*{Appendix C: Comparison of Different Models for the Superconducting Density of States}

Low-temperature scanning tunneling spectroscopy (STM/STS) measurements were carried out in an Omicron low-temperature STM (junction temperature $T \simeq 4.5$~K) and a Unisoku USM1300 He-3 STM (junction temperature $T \simeq 500$~mK), respectively. Spectroscopic measurements were obtained using standard lock-in techniques, with an AC excitation of amplitude of 1.5~mV at 831 Hz for measurements of the normal-state electronic structure, and $60~\mu$V at 931 Hz for measurements in the superconducting state. In all measurements, we fix the current set point 200~pA at 10~mV, ahead of spectroscopy. All spectra of superconducting states through the paper have been spatially averaged, symmetrised about zero bias, and normalised with respect to the normal state conductance, unless otherwise specified.

In the limit of low bias and low temperature ($k_BT \sim eV$), the tunneling differential conductance, measured by scanning probe spectroscopy, is usually expressed as a convolution of the tunneling density of states $N(E)$ and the derivative of the Fermi-Dirac distribution, 

\begin{dmath}\label{eq:LDOS}
    dI/dV \propto \int_{-\infty}^{+\infty} \frac{df(E-eV)}{dE} N(E) dE.
\end{dmath}

For a bulk superconductor with a single isotropic band, a Bardeen-Cooper-Schrieffer (BCS) density of states $N(E) = E / \sqrt{E^2 - \Delta^2}$ is often assumed, and can result in acceptable fits at moderate temperature $T \lesssim T_C$. However, as the derivative of the Fermi-Dirac distribution is a peaked function with half-width $\sim 3.5~k_BT$, it is responsible for thermal smearing of the measured superconducting tunneling density of states. An additional phenomenological broadening parameter, $\gamma$ (Dynes parameter \cite{dynes_1978_function_direct}) is often considered for tunneling experiments, accounting for additional pair-breaking mechanisms.

\begin{equation}
    N(E) = \operatorname{Re}\left[ \frac{(E-i\gamma)}{\sqrt{(E-i\gamma)^2 - \Delta^2}} \right]
\end{equation}

\begin{figure*}
\centering
\includegraphics[clip,width=18cm]{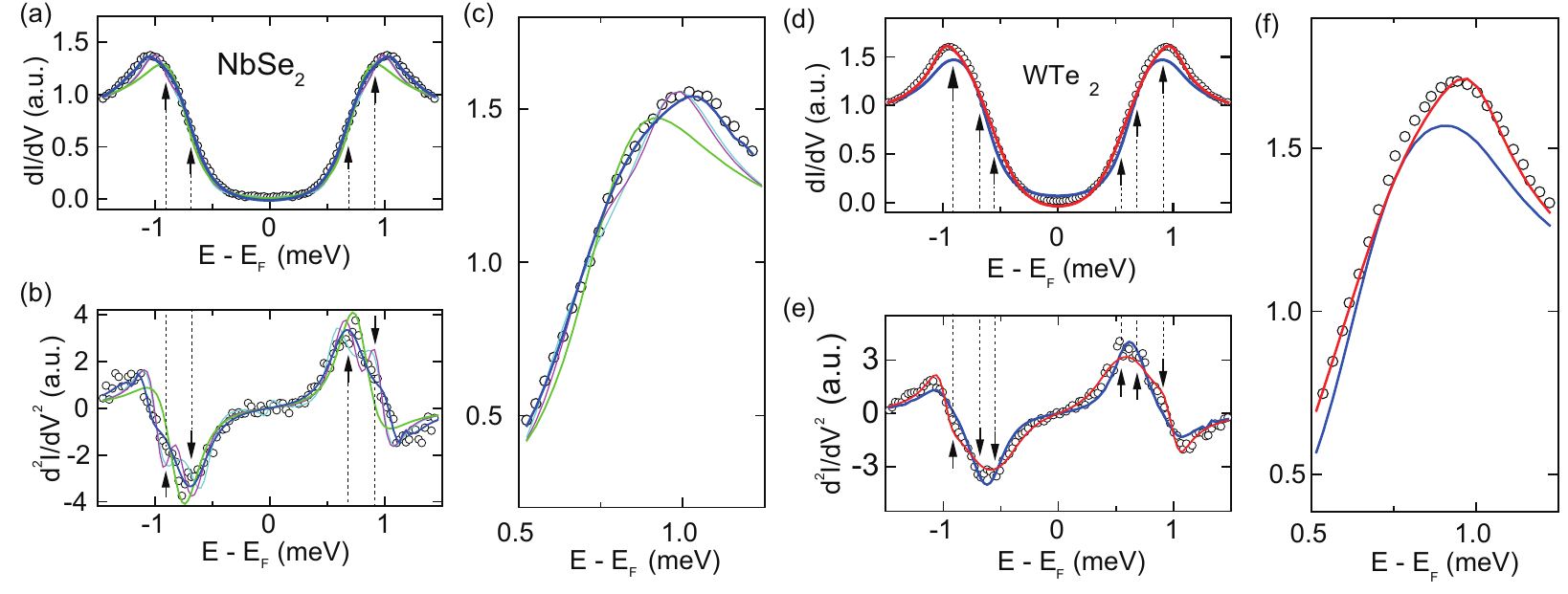}
\caption{(Color online) High-resolution STS spectra of the NbSe$_2$ and WTe$_2$ superconducting gaps at 500~mK, comparing different models. (a) Superconducting energy gap of NbSe$_2$. Solid lines are fits, comparing a single-band BCS/Dynes model (green line), a single anisotropic band model (cyan line), a two-band BCS/Dynes model (magenta line), and a two-band McMillan model (solid blue line). (b), (c) corresponding second derivative of the tunneling current ($d^2I/dV^2$) and detail of the coherence peaks. (d)-(f) Superconducting energy gap of WTe$_2$, comparing a two-band McMillan model (solid blue line) and a three-band McMillan model (solid red line). (e)-(f) same as (b), (c). We observe that a two-band McMillan model is insufficient to describe induced superconductivity in WTe$_2$/NbSe$_2$. All spectra shown have been spatially averaged, symmetrised about zero bias, and normalised with respect to the normal state conductance. Vertical arrows and dashed lines indicate the renormalised/induced gaps, as determined from the two-band and three-band McMillan models, respectively. Across all models, we find Dynes parameters ranging from $0.02 - 0.1$~meV.}
\label{fig:models}
\end{figure*}

At low temperature ($T \ll T_C$), fine detail of the functional form of the superconducting energy gap can be observed \cite{noat_2010signatures, dvir_2018spectroscopy, khestanova_2018unusual}. Indeed, in Fig.~\ref{fig1}(h) we see that a BCS/Dynes model provides a good fit only at $T \simeq 4.5$~K, but fails to describe the detail of the measured superconducting density of states at lower temperature. A more detailed model is needed to describe the low-temperature data, taking the multi-band nature of superconductivity in NbSe$_2$ into account.

In the presence of multiple (non-interacting) bands $i$ at the Fermi energy, the total tunneling density of states may be expressed as a sum over two or more partial densities of states, 
\begin{equation}\label{totalDOS}
    N(E) = \sum\limits_{i} N_i(E)
\end{equation} 
The partial densities of states can further be expressed as
\begin{equation}
    N_i(E) = \tilde{N_i}(E_F) \int_{-\infty}^{\infty} \frac{d\theta}{2\pi} \operatorname{Re} \left[\frac{\lvert{E - i\gamma_i}\rvert}{\sqrt{(E - i\gamma_i)^2 - \Delta_i(E,\theta)^2}}\right]
\end{equation}
where it is assumed that the order parameter $\Delta(E,\theta)$ may not be constant, but itself be a complex energy-dependent and (potentially anisotropic \cite{khestanova_2018unusual}) function. 
The prefactor, $\tilde{N_i}(E_F) = N_i(E_F)T_i$, can be interpreted \cite{noat_2015_2band} as an effective partial density of states weight, renormalized from the true density of states $N_i(E_F)$ by the tunneling probability $T_i$ into band $i$. 

Figure~\ref{fig:models} shows a comparison of different models to fit our data, based on Eq.~(\ref{eq:Ni}), including a single-band BCS/Dynes model ($i=L, \Delta_L$), a two-band BCS/Dynes model ($i = L,S, \Delta_{L,S}$), and an anisotropic single-band model ($i=1$). The second derivative of the tunneling current and close-ups of the coherence peaks are shown alongside to clarify differences in the models. For the anisotropic model we have assumed an empirical anisotropy function, as previously employed for NbSe$_2$ \cite{khestanova_2018unusual}, reflecting the sixfold symmetry of the NbSe$_2$ Fermi surface,
\begin{equation}\label{eq:Khestano}
\Delta(A,\theta) = \Delta_0 (A \ \text{cos}(6 \theta) + (1-A))
\end{equation}

Among the different models, only the two-band BCS and the anisotropic band model provide a somewhat satisfying fit to the measured NbSe$_2$ data \cite{khestanova_2018unusual}. All models, including the anisotropic band fit underestimate the width of the coherence peaks (Fig.~\ref{fig:models}(c)).

\section*{Appendix D: The McMillan Model}

\begin{table*}
\caption{Consistency across two-band and three-band McMillan models, when describing NbSe$_2$ substrate and WTe$_2$/NbSe$_2$ heterostructures, respectively.} 
\begin{center}
    \fontsize{9.5}{12}\selectfont 
    \begin{tabular}{|c||c|c|c|c|c|c|}
    \hline
 Band $i$ & Model &\thead{$\Delta_i^0$ \\ (meV)}&\thead{$\Delta_i({E_F})$ \\ (meV)} & $\frac{N_{S}}{N_{L}}$ & $\frac{N_{\rm WTe_2}}{N_{L}}$ & $\tilde{N}_i(E_F)$\\
\hline\hline
$i=L$&3 band&$1.13\pm0.05$&$0.87\pm0.05$& & &$0.002\pm0.001$\\
\cline{2-7}
(Nb $K$-point)&2 band&$1.11 \pm 0.05$&$1.00 \pm 0.05$& & & \\
\cline{1-7}
$i$=S&3 band&$0.43\pm0.08$&$0.68\pm0.08$&$0.36\pm0.03$& &$0.30\pm0.08$\\ \cline{2-7}
 (Nb $\Gamma$-point)&2 band& $0.45\pm0.09$&$0.73\pm0.09$&$0.36\pm0.01$& & \\ 
\hline
$i$=3&3 band& 0 &$0.58\pm0.02$& &$0.19\pm0.01$&$0.69\pm0.06$\\ 
\cline{2-7}
(WTe$_2$) & 2 band & & & & & \\
\hline
\end{tabular}
\end{center}
\label{table-McM fit-WTe2} 
\end{table*}

Taking the multi-band nature of superconductivity in NbSe$_2$ into account, a compelling model to describe the superconducting density of states was employed by Noat \emph{et al.} \cite{noat_2015_2band}, based on the McMillan equations \cite{mcmillan_1968tunneling}. The McMillan model was originally developed to describe superconductivity between two proximity-coupled materials $i = 1,2$, in which the intrinsic order parameters $\Delta_i^0$ are renormalised by an inter-layer electronic coupling of strength $\Gamma_{ij}$. In the context of multi-band superconductivity \cite{Japanese_2004_MgB2, noat_2010signatures, noat_2015_2band} the inter-layer coupling has since been interpreted as an inter-band scattering rate from band $i$ to band $j$ (``proximity-effect in reciprocal space''). Different from the other models discussed above, the order parameter $\Delta_i(E)$ in the McMillan model is a complex, energy-dependent gap function

\begin{equation}\label{eq:delta}
   \Delta_{i}(E)=\frac{\Delta_i^{0}+\Gamma_{ij}\Delta_j(E)/\sqrt{\Delta_j(E)^2-(E-i\gamma_j)^2}}{1+\Gamma_{ij}/\sqrt{\Delta_j(E)^2-(E-i\gamma_j)^2}}
\end{equation}

Inclusion of these order parameters in Eq.~(\ref{eq:Ni}) gives rise to a set of self-consistent equations to describe the multi-band system, that can be solved numerically to fit the measured tunneling spectra. 

\begin{figure}
\centering
\includegraphics[clip,width=\columnwidth]{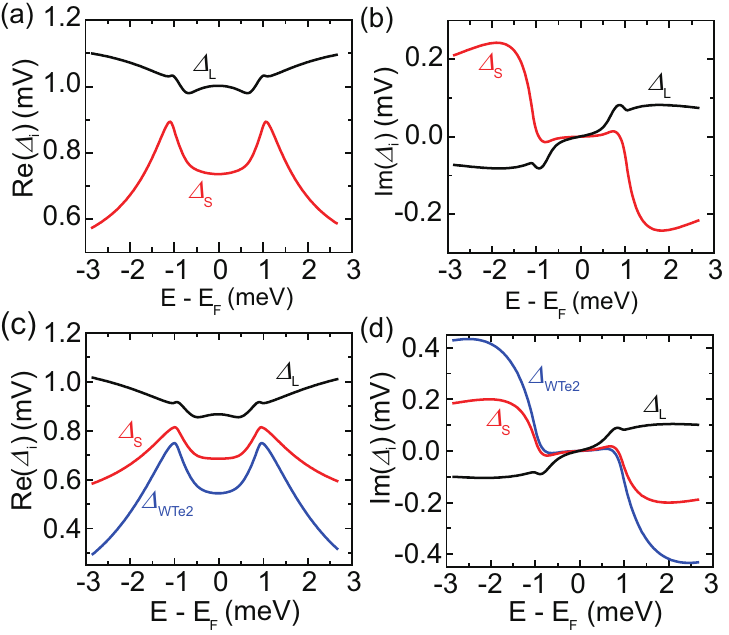}
\caption{(Color online) Energy-dependent order parameters of NbSe$_2$ and monolayer 1T'-WTe$_2$. (a), (b) Real (a) and imaginary (b) parts of the renormalised NbSe$_2$ order parameters, extracted from a two-band McMillan model. (c), (d) Order parameters of 1T'-WTe$_2$/NbSe$_2$, extracted from our three-band McMillan model. $\Delta_{\rm WTe_2}$ denotes the induced order parameter.}
\label{fig:ReIm}
\end{figure}

As highlighted in Ref.~\cite{noat_2015_2band}, in addition to the directly measured partial density of states (Eq.~(\ref{eq:Ni})), and independent estimate of the partial densities of states ratio at the Fermi-energy $\frac{N_j(E_F)}{N_i(E_F)} = \frac{\Gamma_{ij}}{\Gamma_{ji}}$ can be obtained from a ratio of the inter-band coupling parameters. Simultaneous extraction of both during the self-consistent numerical fits thus allow for an accurate extraction. 

To describe multi-band superconductivity in the hybrid WTe$_2$/NbSe$_2$ heterostructure, we consider a third order parameter, $i=3$, in Eq.~\ref{eq:Ni}. To limit the number of free parameters, we consider coupling $\Gamma_{13(31)}$ only between WTe$_2$ and the Nb $K$-band (large gap $\Delta_L^0$), which is expected to have the strongest contribution to the proximity coupling \cite{noat_2015_2band}, and fix $\Gamma_{23(32)} = 0$. We further assume that WTe$_2$ does not carry any intrinsic superconductivity, such that $\Delta_{\rm WTe_2}^0 = 0$. Finally, all NbSe$_2$ related parameters -- determined independently from two-band fits to the substrate -- can be fixed in the three-band model, in particular, $\Delta_{L,S}^{0}$ and $\Gamma_{12(21)}$, leaving only three independent fitting parameters in the three-band model.  

Figures~\ref{fig:models}(d)-\ref{fig:models}(f) shows a comparison our three band McMillan model (solid red line) and the previous two-band model (solid blue line), applied to the WTe$_2$/NbSe$_2$ heterostructure. Importantly, a two-band model cannot represent the data well. From the three band model of the heterostructure, we are able to extract the intrinsic NbSe$_2$ order parameters $\Delta_L^0 = 1.13 \pm 0.05$~meV and $\Delta_S^0 = 0.43 \pm 0.08$~meV, in excellent agreement those extracted from the two-band model of NbSe$_2$ \cite{dvir_2018spectroscopy, noat_2015_2band}. Consistency of the NbSe$_2$ order parameter, extracted separately from two-band and three band models, confirm that the properties of the NbSe$_2$ parent superconductor remain unaffected by the presence of the WTe$_2$ epilayer. The renormalised complex order parameters extracted from the three-band fits are plotted in Figs.~\ref{fig:ReIm}(c) and \ref{fig:ReIm}(d). Here, we observe only a minor re-balancing of the inter-band tunneling rates $\Gamma_{ij}$ with a concomitant small decrease in the Re($\Delta_{L,S}(E_F)$), as a result of the additional coupling to the third band. The value of the induced order parameter, $\Delta_{\rm WTe_2}(E_F) = 0.58 \pm 0.02$~meV agrees well with recent local probe \cite{Hunt_2020proximity} and transport spectroscopy \cite{SC_proxi_Huang_2018inducing} of non-epitaxial heterostructures.

\section*{Appendix F: Mean Field Theoretical Modelling of the Hybrid Bandstructure}

The 1T'-WTe$_2$/2H-NbSe$_2$ heterostructure was modelled using a real-space mean field tight-binding Hamiltonian that is solved using a Bogoliubov-De-Gennes (BdG) formalism. The Hamiltonian contains four separate parts given by, 
	
	\begin{eqnarray}
	H&=&H^{0}_{N}+H_{SC-N}+H^{0}_{W}+H_{W-N} 
	\end{eqnarray}

The individual terms of the Hamiltonian are given by,
	
	\begin{eqnarray}
	H^{N}_{0}&=&\sum_{iljl'\sigma}t^{ll'}_{ij}c^{\dag}_{il\sigma}c^{\dag}_{jl'\sigma} + h.c \nonumber\\
	H^{N}_{SC}&=&  \sum_{il}\Delta^{N}_{il} c^{\dag}_{il\uparrow}c^{\dag}_{il\downarrow}+ h.c\nonumber\\
	H_{W-N}&=&t_{\perp} \sum_{ij\sigma} c^{\dag}_{i2\sigma}d_{j1\sigma} + h.c \nonumber\\
	H^{W}_0&=&\sum_{\mu,\nu ij\sigma,\sigma'}t^{ij}_{\mu,\nu,\sigma,\sigma'}d^{\dag}_{i\mu,\sigma}d_{j\nu \sigma'} + h.c \nonumber
	\end{eqnarray}
	
In $H^{N}_{0}$ operators $(c^{\dag}_{il\sigma},c_{il\sigma})$ represent the creation and annihilation operators, respectively, at the site $i$ with a single Nb $d_{z^2}$ orbital contributing per site (see below for details). The layer index $l=(1,2)$ corresponds to the two layers of 2H-NbSe$_2$, respectively, taking spin $\sigma=(\uparrow,\downarrow)$ into account. The individual real-space tight-binding hopping matrix elements for 2H-NbSe$_2$ are generated by a basis transformation to orbital basis from a two-band Hamiltonian that has been studied previously to match ARPES \cite{rahn2012gaps} and STM experiments \cite{gao2018atomic} on 2H-NbSe$_2$. 

At temperatures below $T\sim 7.2$~K, pristine NbSe$_2$ undergoes a transition to a superconducting state. Assuming a dominant conventional $s$-wave superconducting instability, we calculate the superconducting gap $\Delta^{N}_{il}=V\langle c_{il\uparrow}c_{il\downarrow}\rangle $ self-consistently at each lattice site by solving the BdG equations. The dominant on-site pairing term $V$ is tuned to generate a superconducting gap of $\Delta \sim 1.1$~eV in agreement with the experimental results on pristine 2H-NbSe$_2$ (see Table~\ref{table-NbSe2} and references therein).
 
The electronic structure of 1T'-WTe$_2$ is modelled by generating a real space version of an 8-band Hamiltonian, containing 2 Te and 2 W atoms in a rectangular unit cell. As discussed in Ref~[\onlinecite{lau19,Choe16,Muechler16}], this Hamiltonian corresponds to a dominant contribution from Te $p_x$ orbitals, and W $d_{x^2-y^2}$ orbitals, which dominate the low-energy band structure owing to the distorted lattice structure in monolayer 1T'-WTe$_2$.

In $H^{W}_0$, the indices $i,j$ refer to the unit cell, and $\mu,\nu=(1,2,3,4)$ represent the atoms or orbitals within each unit cell. The model provides reasonable agreement with the low energy electronic structure including the QSHI observed in ARPES and STM experiments \cite{tang_2017quantum}. In the enclosed additional supplementary files, we provide the real space hopping matrix elements including the Rashba spin orbit coupling terms included in the real space Hamiltonian for 1T'-WTe$_2$. The monolayer edge was modelled using an open boundary condition and has been studied for various terminations for directions perpendicular and parallel to the direction of the atomic chains. We find that although the modelling of the edge with open boundary conditions is simplistic (the edge in as as-grown real material is rough with no consistent termination), the obtained local density of states shows reasonable agreement with the experiments (see Fig.~\ref{fig2} of the main text). 

To represent a finite electronic interlayer coupling in the  heterostructure, we assume a nearest neighbor hopping $t_{\perp}$, leading to an effective hybridization between the Nb d$_{z^2}$ orbitals and Te p$_x$ orbitals. This gives rise to a residual metallic density of states within the WTe$_2$ bulk band gap, in reasonable agreement with our experimental observations in both the normal state (see Fig.~2 of the main text) and the superconducting state (see Fig.~3 of the main text). We find best agreement with the experiments -- simultaneously for normal and superconducting state -- for $t_{\perp}\sim 0.15$~eV. 

The induced superconducting gap on the Te and W atoms have been evaluated self consistently by calculating the anomalous averages $\Delta^{W}_{i\mu}\sim \langle d^{\dag}_{i\mu \uparrow}d^{\dag}_{i\mu \downarrow}\rangle $, where $\mu=(1,2,3,4)$ represents the 2 Te and 2 W atoms in the unit cell and $i$ is the unit cell index. The calculations involve a Hamiltonian where the electron operators are considered in momentum space for a direction perpendicular to the edge. For an edge along the $y$-direction, the Bogoliubov-de-Gennes transformations involve quasi-article operators,

\begin{eqnarray}
	d_{i_xk_y\mu\sigma}&=&\sum_{n}(u^n_{i_xk_y\mu\sigma}\gamma_{n\sigma}-\sigma v^{n\star}_{i_xk_y\mu\sigma}\gamma^\dag_{n\sigma})\\
	d^\dag_{i_xk_y\mu\sigma}&=&\sum_{n}(u^{n\star}_{i_xk_y\mu\sigma}\gamma_{n\sigma}^\dag - \sigma v^n_{i_xk_y\mu\sigma}\gamma_{n\sigma})
\end{eqnarray}

Here, $\gamma_{n\sigma}$ are the quasiparticle operators corresponding to state $n$ and $u, v$ are the corresponding amplitudes. Similar transformations also hold for the $c$ operators for NbSe$_2$.

Diagonalizing the above BdG Hamiltonian, we self-consistently calculate the electron density $n(i_x)$ and superconducting gap at each lattice site for the multi-orbital Hamiltonian. The mean-fields for the electron density would in general be given by,

	\begin{eqnarray}
	n_{i_x\mu\sigma} = \frac{1}{N_{k_y}}\sum_{k_y,n}|u^n_{i_xk_y\mu\sigma}|^2 f(E_n)
	\end{eqnarray} 
	Here, $f(E_n)$ is the Fermi function.\\
	
As discussed above, the superconducting gap on NbSe$_2$ has been introduced with a pairing interaction term $V$. The self-consistent procedure leads to an induced even parity ($\Delta^{s}_{i_xj_x\mu}$) order parameter on 1T'-WTe$_2$. The induced superconducting gaps are obtained from the self-consistent solutions by calculating the following anomalous averages,
	
\begin{widetext}
	\begin{eqnarray}
	\Delta^{s}_{i_xj_x\mu}=\frac{1}{N_{k_y}}\sum_{n,k_y=0}^{2\pi}[u^n_{i_xk_y\mu\uparrow}v^{n\star}_{j_xk_y\mu\downarrow}+u^n_{j_xk_y\mu\uparrow}v^{n\star}_{i_xk_y\mu\downarrow}]f(E_n)\nonumber
	\end{eqnarray}
\end{widetext}
	
Here, $N_{k_y}$ is the number of $k_y$ divisions which is typically taken to be $30000-60000$ points to achieve high resolution for the small gaps observed in the system. 

We see our model as a generalization from the approximation of a superconducting heterostructure by considering pristine WTe$_2$ with no effect of the proximal superconductor on the WTe$_2$ band structure, i.e. a heterostructure in which the only effect of the superconductor is to induce a gap on the QSH edges. Instead, our modelling accounts for a realistic finite interlayer coupling between NbSe$_2$ and WTe$_2$ through a self-consistent procedure, allowing for minor changes in the electronic structure such as hybridization and charge transfer. Our model is thus able to explain, simultaneously, a small but finite metallic density of states within the QSH gap, that leaves the edge state intact, and an induced superconducting order parameter of the same magnitude as resolved in the experiments.


\bibliography{PRBbib3}

\end{document}